\documentclass{ws-procs9x6}
\usepackage{graphicx}
\usepackage{latexsym}
\begin{document}

\title{
Radiative Corrections to Neutrino Reactions off Proton and Deuteron
\footnote{\uppercase{P}resented by \uppercase{T.K.} \uppercase{T}he work 
is supported in part by \uppercase{G}rants in 
\uppercase{A}id of the \uppercase{M}inistry of \uppercase{E}ducation. }
}

\author{T. KUBOTA}

\address{Graduate School of Science, Osaka University
\\ 
Toyonaka, Osaka 560-0043, Japan}

\author{M. FUKUGITA}

\address{
Institute for Cosmic Ray Research, University of Tokyo
\\
Kashiwa 277-8582, Japan}

\maketitle

\abstracts{
Radiative corrections are calculated for antineutrino proton
quasielastic scattering, neutrino deuteron scattering, and
the asymmetry of polarised neutron beta decay from which
$G_{A}/G_{V}$ is determined. A particular emphasis is given to the constant
parts that are usually absorbed into the coupling constants, and thereby
those that appear in the processes that concern us are unambiguously tied
among each other.
}

\section{Introduction}
Neutrino experiments have now entered 
the era of precision measurements with
the accuracy reaching the level that 
radiative corrections cannot be ignored.
In this talk 
we  report our recent 
work \cite{fuku1}\cdash \cite{fuku3} 
on the radiative corrections to neutrino scattering,
$\bar \nu _{e}+p\longrightarrow e^{+}+n$ 
as measured in KamLAND and  $\nu _{e} +d \longrightarrow  e^{-}+p+p$,
$\nu _{x} +d \longrightarrow \nu _{x} + p+n$ that are measured at
SNO, 
and those to the asymmetry in polarised neutron beta decay. 

\section{Radiative Corrections to $G_{A}/G_{V}$ }
The study of radiative corrections to weak processes has a long 
history, 
and the wisdom acquired for neutron 
beta decay rate\cite{kinoshita}\cdash \cite{sirlin2} 
 will be transcribed in the calculation of neutrino scattering processes. 
With the local four-Fermi interactions, the $G_{A}/G_{V}$ 
ratio ($=g_A$) enters  
the tree-level cross sections as
\begin{equation}
\sigma (\bar \nu _{e}+p\longrightarrow e^{+}+n) \propto 
(f_{V}^{2}+3g_{A}^{2})
\label{eq:kamland}
\end{equation}
for neutrino-proton quasielastic scattering and 
\begin{eqnarray}
\sigma (\nu _{e} +d &\longrightarrow & e^{-}+p+p) \propto 
g_{A}^{2}, \hskip3.4cm ({\rm CC})
\label{eq:cc}
\\
\sigma (\nu _{x} +d &\longrightarrow & \nu _{x} + p+n) \propto 
g_{A}^{2}
\hskip1cm [x=e, \mu , \tau] \hskip0.8cm ({\rm NC})
\label{eq:nc}
\end{eqnarray}
for scattering off deuteron.
Here $f_{V}=1$  
is retained to trace the contribution of  the weak vector 
current.

We separate the radiative corrections to charged current processes  
into the inner and outer parts as was done for neutron 
beta decay. The outer corrections depend on 
the  $e^\pm$ 
velocity  in the final state and are 
free from UV and IR divergences. They are independent 
of the details of strong interactions and the calculation 
is straightforward. The inner part  
is velocity-independent but is plagued by the 
UV-divergence that is not cancelled within four-Fermi theory and 
depends on the details of the structure of nucleons.
The inner corrections amount to replacing  
$f_{V}^{2}$ and $g_{A}^{2}$ by
\begin{eqnarray}
\bar f_{V}^{2}=f_{V}^{2}(1+\delta _{\rm in}^{{\rm F}}), \hskip0.5cm
\bar g_{A}^{2}=g_{A}^{2}(1+\delta _{\rm in}^{{\rm GT}}). 
\label{eq:innercorrections}
\end{eqnarray}
The evaluation of  $\delta _{\rm in}^{\rm F}$ and 
$\delta _{\rm in}^{\rm GT}$ requires not only renormalisable 
Weinberg-Salam theory but receives complications from
hadron structure. 
For the Fermi transitions 
$\delta _{\rm in}^{\rm F}$ 
is already obtained\cite{marciano}\cdash\cite{towner}, 
but the evaluation of 
$\delta _{\rm in}^{\rm GT}$ for the Gamow-Teller transitions 
has eluded the literature for long time. 

The corrections to (\ref{eq:nc}) differ from those to 
(\ref{eq:kamland}) and 
(\ref{eq:cc}).  There is no outer correction. The 
corrections are incorporated  by the replacement of $g_{A}^{2}$  by 
\begin{eqnarray}
g_{A}^{2} \longrightarrow g_{A}^{2}(1+\Delta _{\rm in}^{\rm GT}).
\label{eq:ncreplacement}
\end{eqnarray} 

The prime purpose of our work is to give 
$\delta _{\rm in}^{\rm GT}$ and $\Delta _{\rm in}^{\rm GT}$,
so that the $g_{A}$ factor that appears in 
NC processes is related with that in the CC processes.
Subsidiarily we show that the radiative correction to the polarised 
neutron beta decay asymmetry (from which we determine $g_A$) 
is described by the
same factors as those that appear in (\ref{eq:innercorrections})
and in the neutron beta decay rate.
This is expected, but we do not find any proofs. So we carried out
explicit calculations.
The $G_{A}/G_{V}$ ratio measured in this process is 
\begin{eqnarray}
\frac{g_{A}\left (1+\delta _{\rm in}^{\rm GT}\right )^{1/2}}
{f_{V}\left (1+\delta _{\rm in}^{\rm F}\right )^{1/2}}\approx g_{A}
\left (1+\frac{\delta _{\rm in}^{\rm GT}-\delta _{\rm in}^{\rm F}}{2}\right )
\ne g_A\ . 
\label{eq:ga/gvratio}
\end{eqnarray}

\section{Calculational Strategy}
\label{sec:strategy}

Let us begin with the charged current processes (\ref{eq:kamland}) and 
(\ref{eq:cc}).
Following the procedure for the neutron beta decay 
rate\cite{sirlin2}, we divide the integration region of the exchanged gauge 
boson into 
\begin{eqnarray}
({\rm i}) \:\: 0<\vert k \vert ^{2}<M^{2}, \hskip0.5cm {\rm and}\hskip0.5cm
({\rm ii}) \:\: M^{2}<\vert k\vert ^{2}< \infty . 
\label{eq:iandii}
\end{eqnarray}
The mass scale $M$ 
is supposed to be greater than the proton mass ($m_{p}$), 
but to be much smaller than the $W$ and $Z$ boson masses 
($m_{W}$, $m_{Z}$), {\it i.e.,} 
$m_{p} \ll M \ll m_{W}, m_{Z}$.
We use the four-Fermi interactions for nucleons
in region (i), thereby dealing with 
the nucleons as point-like and only photons are exchanged between 
nucleons and charged leptons. 
In region (ii)  we employ 
Weinberg-Salam theory for quarks and photons and $Z$ bosons are exchanged. 
The mass scale $M$ is the UV-cutoff for the four-Fermi theory,  
but is also regarded as  the scale for the onset of
the asymptotic behaviour to which Weinberg-Salam theory applies.

With four-Fermi theory, we end up with UV-divergences, {\it i.e.,} 
${\rm log}M^{2}$ terms in the calculation for (i). 
These divergences are classified into two types, the one
eliminated by renormalisable gauge  theories and the 
other that is 
rendered finite only by considering the structure of nucleons. 
It is known \cite{abers}\cdash \cite{sirlin} that 
such classification is possible for the Fermi 
transitions by using CVC and current algebra techniques. 
We showed in Ref. \refcite{fuku1} 
that a parallel classification is possible 
for the Gamow-Teller transitions on the basis of 
CVC, PCAC and current algebra (see also Ref. \refcite{sirlin3}).

In the first type ${\rm log}M^{2}$ terms are universal, 
their coefficients being independent of the details of strong interactions. 
These ${\rm log}M^{2}$ terms are cancelled when the integrals in
(i) and (ii) are added. 
The ${\rm log}M^{2}$ terms that appear axial-vector vector interference
terms cannot be cancelled. So,  
 ${\rm log}M^{2}$ terms in (i) are tamed  by introducing form factors 
 at the electromagnetic and weak vertices in Feynman integrals. 
Weak magnetism cannot be ignored\cite{fuku1} at the weak vertices, because 
the mass scale of the form factor is on the order of $m_{p}$ and the loop 
integral over the weak magnetism form factor gives the same order 
of magnitude as does the $V-A$ contribution.

\section{Antineutrino Quasielastic Scattering off Proton}

We write the differential cross section 
as 
\begin{eqnarray}
\frac{d\sigma (\bar \nu _{e}+p\longrightarrow e^{+}+n)}
{d({\rm cos}\theta )}=\frac{G_{V}^{2}}{2\pi }E^{2}\beta \left \{
A(\beta )+B(\beta ) \beta {\rm cos}\theta 
\right \}. 
\label{eq:diffcrossection}
\end{eqnarray}
Here $E$ and $\beta $ are the energy and velocity of the final positron, 
$\theta $ is the angle between incident antineutrino and the positron and 
$G_{V}=G_{F}{\rm cos}\theta _{C}$ is the vector coupling constant to
nucleons.

After calculations we find that $A(\beta )$ and $B(\beta )$ are written
\begin{eqnarray}
A(\beta )&=&\left \{  1+\delta _{\rm out} (E) \right \}
\left ( \bar f_{V}^{2}+3\bar g_{A} ^{2} \right ), 
\\
B(\beta )&=&\left \{  1+ \tilde \delta _{\rm out} (E) \right \}
\left ( \bar f_{V}^{2}-\bar g_{A} ^{2} \right ),
\label{eq:AB}
\end{eqnarray}
where the inner corrections included in $\bar f_{V}^{2}$ and 
$\bar g_{A}^{2}$ are given in (\ref{eq:innercorrections})
with
 \begin{eqnarray}
\delta _{\rm in}^{\rm F}&=&\frac{e^{2}}{8\pi ^{2}}\left \{
4{\rm log}\left ( \frac{m_{Z}}{m_{p}} \right ) +{\rm log}
\left (\frac{m_{p}}{M} \right )+C^{\rm F}
\right \},
\label{eq:deltainF}
\\
\delta _{\rm in}^{\rm GT}&=&\frac{e^{2}}{8\pi ^{2}}\left \{
4{\rm log}\left ( \frac{m_{Z}}{m_{p}} \right ) +{\rm log}
\left (\frac{m_{p}}{M} \right )+1+C^{\rm GT}
\right \}.
\label{eq:deltainGT}
\end{eqnarray}
The effect of the nucleons structure appears  only in 
$C^{\rm F}$ and $C^{\rm GT}$. We computed 
these two numbers by introducing  
form factors. The contributions from the 
weak magnetism are nonnegligible at the weak vertex; it even dominates 
$C^{\rm GT}$. 
Our results are $C^{F}=2.160$ and $C^{\rm GT}=3.281$, 
and hence $\delta _{\rm in}^{\rm F}=0.0237$ and 
$\delta _{\rm in}^{\rm GT}=0.0262$ for $M\approx 1$ GeV .

One sees in (\ref{eq:AB}) that the energy-dependent outer corrections 
are factored out.
One of them, $\delta _{\rm out}(E)$, has been known\cite{vogelfayans}; 
the other 
$\tilde \delta _{\rm out}(E)$ 
is new\cite{fuku1}. 
The outer corrections for (\ref{eq:cc}) are given in Ref.~\refcite{towner2}.

\section{Asymmetry in Polarised Neutron Beta Decay}

The $G_{A}/G_{V}$ ratio is determined from
the asymmetry parameter $A$. 
The corrections are again separated into the inner and outer parts, 
as\cite{fuku2} 
\begin{eqnarray}
A=2\left \{ 1+ C(E) \right \}
\frac{\bar f_{V}\bar g_{A}-\bar g_{A}^{2}}{\bar f_{V}^{2}+3\bar g_{A}^{2}}. 
\label{eq:asymmetry}
\end{eqnarray}
The energy-dependent factor  $C(E)$ is given in Ref. \refcite{fuku2}. 
The important point is that exactly the same inner corrections appear in 
$\bar f_{V}^{2}$ and $\bar g_{A}^{2}$, as given  by   
(\ref{eq:innercorrections}) with 
(\ref{eq:deltainF}) and (\ref{eq:deltainGT}), 
so that  $\bar g_{A}$ from asymmetry can be used to predict the
neuron decay rate without further corrections.

\section{The Deuteron Neutral Current Process}

Since the outer correction is absent,
 all we need is the evaluation of the 
inner part using Weinberg-Salam theory on the quark level. 
This type of calculation was done by Marciano and Sirlin\cite{marcianosirlin}.
 The effective interaction of quarks and neutrinos 
at low energy with an iso-singlet target is  given  by 
\begin{eqnarray}
M_{\rm eff}&=&-i \frac{G_{F}}{\sqrt{2}}
\rho _{\rm NC}^{(\nu;h)}\:
\bar \psi _{\nu}\gamma ^{\mu}(1-\gamma ^{5})\psi _{\nu}
\nonumber \\ & & \times 
\left \{
\bar \psi I_{3}\gamma _{\mu}(1-\gamma ^{5})\psi -2\: 
\kappa ^{(\nu; h) }\:
{\rm sin}^{2}\theta _{W}\bar \psi  \gamma _{\mu}Q \psi
\right \}, 
\label{eq:effective1}
\end{eqnarray}
where $\psi _{\nu}$ and  $\psi $ are the neutrino and  the quark doublet, 
and $\rho _{\rm NC}^{(\nu;h)}-1$
and $\kappa ^{(\nu; h) }-1$ are the radiative corrections, which are 
found in Ref. \refcite{marcianosirlin}.

If we sandwich (\ref{eq:effective1}) between the deuteron and two-nucleon 
states to evaluate the cross section  (\ref{eq:nc}), 
only the axial current in (\ref{eq:effective1}) contributes to the 
${}^{3}S_{1}\longrightarrow {}^{1}S_{0}$ transition.
The effective interaction for the 
nucleon doublet $\psi _{N}$  reads
\begin{eqnarray}
M_{\rm eff}=i \frac{G_{F}}{\sqrt{2}}\: 
g_{A}\:\rho _{\rm NC}^{(\nu; h)}
 [ \bar \psi  _{\nu}\gamma ^{\mu}(1-\gamma ^{5})\psi _{\nu}
 ] [ \bar \psi _{N} I_{3} \gamma _{\mu}\gamma ^{5}\psi _{N} ].
 \label{eq:effective2} 
\end{eqnarray}
The term of $\kappa ^{(\nu; h)}$ does not contribute 
to (\ref{eq:nc}).
Eq. (\ref{eq:effective2}) indicates that $g_{A}$ is renormalised 
multiplicatively by $\rho _{\rm NC}^{(\nu; h)}$. 
Writing 
$\rho _{\rm NC}^{(\nu;h)}=\left ( 1+\Delta _{\rm in}^{\rm GT} \right )^{1/2}$, 
the radiative correction is the replacement (\ref{eq:ncreplacement}).
We compute 
$\rho _{\rm NC}^{(\nu ; h)}=1.00955$  for  Higgs boson mass
$m_{H}=1.5 m_{Z}$
and 
1.00862  for  $m_{H}=5 m_{Z}$.

\section{Summary}

We computed radiative corrections to antineutrino proton
quasielastic scattering, neutrino deuteron scattering, and
the asymmetry of polarised neutron beta decay, with an emphasis 
given to the constant
parts that are usually absorbed into the coupling constants.
Hereby couplings that appear in the processes that concern us 
are unambiguously tied. For instance, the NC to
CC ratio for neutrino-deuteron reactions receives the overall
correction $(1+\Delta _{\rm in}^{\rm GT})/(1+\delta _{\rm in}^{\rm GT})
=0.992\pm0.001$ up to the outer correction for CC which is accounted for
separately.


\end{document}